\documentclass[reprint,amsmath,amssymb,aps,prd,nofootinbib]{revtex4-2}
\usepackage[utf8]{inputenc}
\usepackage{graphicx}
\usepackage{dcolumn}
\usepackage{bm}
\usepackage[most]{tcolorbox} 
\usepackage[colorlinks=true,   
            citecolor=blue,    
            urlcolor=blue,     
            linkcolor=blue]{hyperref}
\usepackage{amsmath}
\usepackage[caption=false]{subfig}
\usepackage{microtype}
\usepackage{placeins}

\def\be{\begin{equation}}
\def\ee{\end{equation}}

\newcommand{\bea}{\begin{eqnarray}}
\newcommand{\eea}{\end{eqnarray}}
\newcommand{\barr}{\begin{array}}
\newcommand{\earr}{\end{array}}

\begin{document}

\title{Quantum Critical Collapse Abhors a Naked Singularity}%

\author{Marija Toma\v{s}evi\'c}
\email{m.tomasevic@uva.nl}
\affiliation{Institute for Theoretical Physics, University of Amsterdam, Science Park 904, 1090 GL Amsterdam, The Netherlands}

\author{Chih-Hung Wu}
\email{chwu29@uw.edu}
\affiliation{Department of Physics, University of Washington, Seattle, WA 98195, USA}

\begin{abstract}

Classical critical collapse yields naked singularities from smooth initial data, challenging cosmic censorship, and shaping the spectrum of primordial black holes. We show that one-loop vacuum polarization near the threshold qualitatively changes this outcome by dressing the singularity with a horizon within a controlled semiclassical regime. In analytically tractable Einstein-scalar critical spacetimes, a one-loop $s$-wave treatment linearized around self-similar backgrounds shows that regularity uniquely selects an asymptotically Minkowskian, vacuum-polarization state. Its renormalized stress tensor carries a universal quantum growing mode that competes with the classical unstable mode, shifts the critical point, and generates a trapped surface along with a finite mass gap at the new threshold, thereby enforcing horizon formation even under arbitrary fine-tuning. In primordial collapse, the threshold shift enters exponentially into the formation fraction, while the mass gap truncates the low-mass tail, suggesting potentially important consequences for the predicted mass spectrum. These results provide a self-consistent semiclassical treatment of critical collapse and yield sharp predictions within the one-loop, near-critical, linearized regime.

\end{abstract}

\maketitle

\section{Introduction}

The fate of spacetime singularities is of central importance in gravity: every classical black hole ends in one, and every expanding universe originates from one. If such singularities are not hidden behind horizons, classical determinism fails, compelling us to confront quantum gravity. At the heart of this conundrum lies the cosmic censorship conjecture~\cite{Penrose:1969pc}; as Hawking remarked, ``\textit{Nature abhors a naked singularity}''~\cite{Hawking:1994sss}.

Explicit counterexamples involving naked singularities are known. The most familiar one is the endpoint of black hole evaporation, where Hawking radiation shrinks the horizon, ultimately revealing a naked singularity~\cite{Hawking:1974sw,Hawking:1976ra}. Another example is the Gregory-Laflamme instability~\cite{Gregory:1993vy}, in which higher-dimensional black strings may classically fragment toward naked pinch-off singularities.
%\footnote{See \cite{Horowitz:1996nw, Horowitz:1997jc, Ceplak:2023afb, Seitz:2025wpc, Emparan:2024mbp} for stringy resolutions of these counterexamples.}

Yet the most physically realistic setting is classical critical gravitational collapse~\cite{PhysRevLett.70.9, Gundlach:2025yjee}, wherein finely tuned smooth initial data drive the formation of naked singularities. This process can arise in the early universe and has been identified as a universal feature in scenarios involving primordial black hole (PBH) formation~\cite{Niemeyer:1997mt}. 

In this paper, we show that quantum effects qualitatively overturn this classical picture already within a controlled semiclassical regime. We study analytically tractable Einstein-scalar critical spacetimes using a one-loop $s$-wave treatment linearized around self-similar near-critical backgrounds. In this setting, regularity in the self-similar domain fixes the relevant asymptotically Minkowskian, vacuum-polarization state without artificial quantum flux, the renormalized stress tensor develops a universal quantum growing mode, and one-loop backreaction competes with the classical unstable mode to shift the critical threshold. The resulting semiclassical geometry develops a trapped surface and a finite mass gap, so that the would-be naked critical endpoint is replaced by horizon formation. The issue we address is therefore not the ultimate Planck-scale fate of the singularity, but whether semiclassical vacuum polarization already changes the threshold dynamics before that regime is reached. Our analysis shows that indeed it does.

\section{Classical critical collapse}

Critical collapse was first discovered by Choptuik~\cite{PhysRevLett.70.9} in the spherically symmetric collapse of a massless scalar field---the prototypical Einstein-scalar system~\cite{Christodoulou:1986zr,Christodoulou:1991yfa,Christodoulou:1993,Christodoulou:1994hg,Christodoulou_1999}. It probes the threshold of black hole formation in the space of initial data. For any one-parameter family $p$ of initial configurations interpolating between dispersion and black hole formation, there exists a critical value $p^\ast$. For marginally supercritical data $p>p^\ast$, the black hole mass obeys the universal scaling law
\be\label{eq:scalinglaw}
M_{\text{BH}} \propto (p-p^\ast)^\gamma,
\ee
where the exponent $\gamma$ depends only on the matter model. In the Einstein-scalar system, $\gamma \simeq 0.37$. As $p\to p^\ast$, the black hole mass tends to zero and the limiting spacetime contains a naked singularity. This scaling law, together with the absence of a mass gap, is the hallmark of so-called Type II critical collapse. By contrast, Type I critical collapse is characterized by a non-zero mass gap; in that case, the scaling law applies instead to the lifetime of the intermediate metastable configuration.

%By analogy with statistical mechanics, $M_{\text{BH}}$ plays the role of an order parameter; Type I and Type II collapse are distinguished by the presence or absence of a mass gap. The Einstein-scalar system exhibits robust Type II critical behavior.

In the high-curvature region preceding black hole formation, the spacetime evolves toward a self-similar critical solution, largely independent of the initial-data family~\cite{Gundlach:1995kd,Gundlach:1996eg,Gundlach_2003,Martin-Garcia:2003xgm,Frolov:2003dk}. Self-similarity means that the solution repeats its structure at progressively smaller scales, either continuously or discretely. Choptuik's scalar critical solution is discretely self-similar (DSS), recurring only after a fixed logarithmic rescaling set by the echoing period.

On the other hand, continuously self-similar (CSS) critical solutions are analytically tractable and admit a homothetic Killing vector
\be
\mathcal{L}_\xi g_{\mu\nu} = 2 g_{\mu\nu}, \qquad \xi=-\frac{\partial}{\partial T},
\ee
allowing the metric to be written as
\be
g_{\mu\nu}(T,x^i)=\ell^2 e^{-2T}\tilde g_{\mu\nu}(x^i),
\ee
where $\ell$ is an arbitrary length scale. In the critical spacetime, curvature invariants diverge as $T\to\infty$, corresponding to the naked singularity. The critical solution acts as a codimension-one attractor in phase space and possesses exactly one unstable spherically symmetric mode~\cite{Gundlach:1995kd,Gundlach:1996eg,Martin-Garcia:1998zqj}. This unique growing mode determines whether the evolution disperses or forms a black hole.

\begin{figure}[htbp]
\centering
\includegraphics[width=0.3\textwidth]{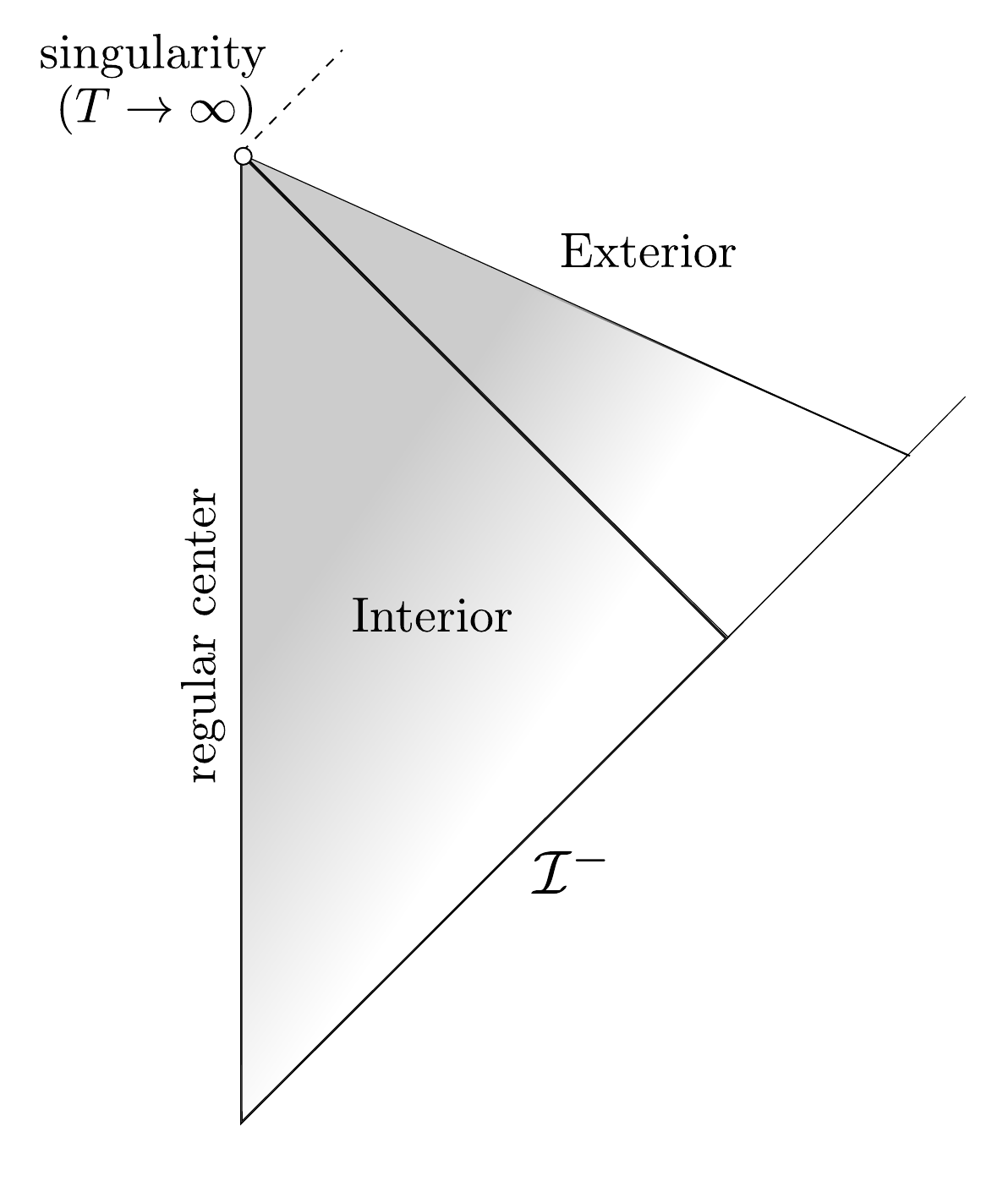}
\caption{Global structure of a CSS critical spacetime, in which the past light cone of the naked singularity naturally separates the spacetime into interior and exterior regions~\cite{Gundlach_2003, Martin-Garcia:2003xgm, Cicortas:2024hpkk}. Our focus is on the interior region, where the singularity dynamically emerges from smooth initial data.}
\label{cssglobal}
\end{figure}
\FloatBarrier

\section{The quantum effective theory}

\paragraph{\textbf{Quantum critical collapse.}} The classical picture is incomplete. In the self-similar phase of Type II collapse, curvature grows without bound on ever smaller scales, precisely where quantum effects should start to play a role. This regime offers a natural arena for incorporating quantum corrections from collapsing matter before a full quantum gravity treatment becomes unavoidable. The question is not whether full quantum gravity ultimately resolves the singularity, but whether semiclassical vacuum polarization already changes the threshold dynamics in a controlled way.

Semiclassical analyses of critical spacetimes have nevertheless remained limited over the past thirty years~\cite{Strominger:1993tt,Peleg:1996ce,Bose:1996pi,Ayal:1997ab,Chiba,Brady:1998fh,Berczi:2021hdh,Moitra:2022umq}, and no consensus has emerged. Existing approaches often assume conformal matter or impose quantum states that introduce spurious Hawking-like flux even before horizon formation. These assumptions obscure the specific dynamical role of vacuum polarization in the near-critical regime and lead to conflicting conclusions~\cite{Tomasevic:2025clf}. Our aim is to isolate the physical semiclassical effect that is actually relevant there.

Since all non-spherical perturbations decay except for a single spherical growing mode, it is sufficient to focus on the dominant $s$-wave sector when describing the dynamics. For a minimally coupled massless scalar field $f$ in $D=d+1$ dimensions, one isolates the $s$-wave via spherical reduction to two dimensions with a dilaton $\phi$~\cite{Grumiller:2002nm},
\be
ds^2_{(D)} = g_{ab}dx^a dx^b + e^{\frac{-4\phi}{D-2}} d\Omega^2_{D-2},
\ee
under which the matter Lagrangian acquires a universal dilaton coupling
\be
\mathcal{L}_{\text{matter}} \propto -\sqrt{-g}\,e^{-2\phi}(\nabla f)^2.
\ee
Although the reduced matter sector is no longer a free conformal theory, it retains two-dimensional Weyl invariance. This makes it possible to compute the one-loop exact trace anomaly via heat-kernel methods in path-integral quantization, independent of the state and regularization scheme. The renormalized trace is~\cite{Mukhanov:1994ax, Bousso:1997cg, Kummer:1998dc, Kummer:1999zy, Balbinot:2000iy, Fabbri:2003vy, Hofmann:2004kk}
\be\label{eq:anomaly}
\langle T^a{}_a\rangle = \frac{\hbar}{24\pi}\left(R-6(\nabla\phi)^2+6\Box\phi\right).
\ee
The one-loop effective action contains anomaly-induced non-local pieces whose variation reproduces Eq.~\eqref{eq:anomaly}:
\be\label{eq:oneloopaction}
\Gamma \sim -\frac{\hbar}{96\pi}\int \left(R\Box^{-1}R -12(\nabla\phi)^2\Box^{-1}R +12\phi R\right).
\ee
Additional Weyl-invariant terms may appear. They do not affect the trace, but they can depend on the state and need not admit simple closed forms~\cite{Karakhanian:1994gs,Jackiw:1995qhh,Navarro-Salas:1995lmi, Buric:1998xv, Balbinot:1998yh, Balbinot:1999ir, Buric:2000cj, Balbinot:2002bz, Hofmann:2004kk, Hofmann:2005yv, Wu:2023uyb, Shi:2026wnk}. In the present problem, however, the regular state singled out by the self-similar background is insensitive to these ambiguities~\cite{Tomasevic:2025clf}.

To make the action local, we introduce classical auxiliary fields $\chi_1$ and $\chi_2$ satisfying
\be\label{eq:auxfields}
\Box\chi_1 = \lambda_1 R + \lambda_2 (\nabla\phi)^2,
\qquad
\Box\chi_2 = -\mu_1 R - \mu_2 (\nabla\phi)^2,
\ee
with coefficients $\{\lambda_i,\mu_i\}$ determined by matching to Eq.~\eqref{eq:oneloopaction}. The homogeneous parts of $\chi_{1,2}$ encode the choice of Green's function and hence the quantum state. Varying the localized one-loop action with respect to the metric yields $\langle T_{ab}\rangle$.

\paragraph{\textbf{Regularity and the choice of state.}}
What fixes the quantum state is not an appeal to a static black-hole vacuum, but regularity within the physical self-similar domain. Concretely, one requires that $\langle T_{ab} \rangle$ be free of divergences on the symmetry axis and on the past light cone of the singularity inside the CSS patch, while remaining asymptotically Minkowskian and carrying no imposed quantum flux. It turns out that these conditions determine the homogeneous sectors in $\chi_{1,2}$ and single out a unique vacuum-polarization state.

This state is \emph{Boulware-like} rather than Unruh-like. In black-hole spacetimes, Unruh-like states are appropriate when one wishes to describe a geometry with a formed future horizon and outgoing Hawking flux at future null infinity. Our problem is different. We are interested in the earliest near-critical, pre-horizon regime and in the first marginally trapped surface, where the relevant quantum effect is the vacuum polarization of the collapsing matter itself, not a prescribed Hawking flux from an already formed black hole. In this sense the state is Boulware-like: it captures genuine vacuum polarization and makes the stress tensor vanish near the asymptotic region.\footnote{Regularity should also not be misunderstood as a claim that the collapsing scalar profile is non-singular all the way to the endpoint $T\to\infty$. In adapted self-similar coordinates, $T$ is a logarithmic scale variable, and the limit $T\to\infty$ corresponds to the approach to the classical naked singularity. Our semiclassical analysis is formulated at finite, moderately large $T$, where one-loop backreaction and linearized perturbation theory remain controlled. On the contrary, we will see that the growth of the $\langle T^{(D)}_{\mu \nu} \rangle$ with $T$ is precisely the physical signal that vacuum polarization becomes increasingly important as the singular regime is approached.} However, once a global exterior and future horizon have formed, an Unruh-like evaporation description may become appropriate for that later global evolution, which lies outside the local pre-horizon analysis used here to locate the first trapped surface.

Compatibility with the conservation law lifts the result back to the full $D$-dimensional stress tensor through the $s$-wave relations~\cite{Mukhanov_1994}
\be\label{eq:swave}
\langle T^{(D)}_{ab}\rangle \propto \frac{\langle T^{(2)}_{ab}\rangle}{e^{-2\phi}},
\qquad
\langle T^{(D)}_{\theta\theta}\rangle \propto e^{\frac{2\phi(D-4)}{D-2}} \frac{1}{\sqrt{-g}}\frac{\delta\Gamma}{\delta\phi}.
\ee

The anomaly-based computation of one-loop backreaction is robust, consistent with Wald's axiomatic framework~\cite{Wald:1977up,Wald1978,Wald:1978ce}, and well-suited to genuinely dynamical self-similar geometries without an exact timelike Killing vector. To demonstrate its power and generality, we now apply it to two analytically tractable CSS Einstein-scalar spacetimes: the Garfinkle solution in $D=2+1$~\cite{Garfinkle:2000br} and the Roberts solution in $D=3+1$~\cite{Roberts:1989sk,Brady:1994xfa,Oshiro:1994hd}. Neither is the exact numerical critical solution of its corresponding collapse problem, but each captures essential features of near-critical self-similar dynamics and allows the semiclassical mechanism to be exhibited in closed form.

\section{Quantum critical spacetimes}

\paragraph{\textbf{Garfinkle spacetime in 2+1 dimensions.}} This family of critical spacetimes is labeled by a positive integer $n$, with the metric:
\begin{equation} \label{eq:Garfinkle}
    ds^2=  e^{-2 T} \bigg[e^{2 \rho_0} \bigg(dx-\frac{x}{2n} dT \bigg)dT+r_0^2 d \theta^2 \bigg],
\end{equation}
where
\begin{equation}
    e^{2\rho_0}=2n \bigg(\frac{1+x^n}{2} \bigg)^{4(1-\frac{1}{2n})}, \quad r_0=\frac{1-x^{2n}}{2},
\end{equation}
with $T \in (- \infty, \infty)$ and $x \in [0,1]$. The scalar field supporting this geometry is (with units $c=1$, $8 \pi G_N=1$)
\begin{equation}
    f(T,x)=\sqrt{\frac{2n-1}{2 n}}\bigg[T-2 \ln{\bigg(\frac{1+x^n}{2} \bigg)}\bigg].
\end{equation}
A curvature singularity appears as $T \to \infty$, with the global structure resembling the interior region of Fig.~\ref{cssglobal}. In particular, the $n=4$ case shows remarkable agreement with numerical simulations~\cite{Jalmuzna:2015hoa}. 

The exact Garfinkle solution is defined in an asymptotically flat spacetime with cosmological constant $\Lambda = 0$. However, the spectrum of perturbations has subtleties when interpreted as a true critical solution, and the detailed numerical agreement with
collapse simulations is obtained only after taking into account the quasi-CSS completion with perturbative $\Lambda <0$
corrections~\cite{Jalmuzna:2015hoa}. %However, we emphasize that the exact Garfinkle solution itself requires the cosmological constant $\Lambda=0$, its perturbation spectrum has subtleties when interpreted as a true critical solution, and the detailed numerical agreement with collapse simulations is obtained only after taking into account the quasi-CSS completion with perturbative $\Lambda <0$ corrections~\cite{Jalmuzna:2015hoa}.  
These features are model-specific and should
not be conflated with universal properties of critical collapse in higher dimensions. In Section~\ref{sec:discussion}, we explain which features extracted from this example are expected to be generic.

At leading self-similar order, however, the quantum shielding mechanism does not rely on $\Lambda$. The quantum growing mode we are about to reveal is already present in the CSS regime. The cosmological constant becomes important only when one connects the self-similar interior to the full physical collapse geometry, so $\Lambda$ can enter the physical normalization of the threshold shift and mass gap through quasi-CSS matching and boundary data.

The analytic structure of this geometry permits closed-form expressions for $\langle T^{(3)}_{\mu \nu}\rangle$ of any integer $n$. Regularity within the self-similar domain $x \in [0, 1]$ fixes the homogeneous parts of the auxiliary fields and uniquely selects a Boulware‐like state described above. The resulting stress tensor takes
the remarkably simple form
\begin{equation} \label{eq:Garfinkletensor}
    \langle T^{(3)}_{\mu \nu} \rangle = e^T F_{\mu \nu}(x,n),
\end{equation}
where $F_{\mu \nu} (x, n)$ is real-analytic in $x$; see~\cite{Tomasevic:2025clf}. The overall $e^T$ growth has a clear physical origin: it reflects the self‐similar rescaling of the areal radius, carried by the dilaton $\phi$ through the $s$-wave reduction in Eq.~\eqref{eq:swave}. One-loop vacuum polarization
therefore generates a genuine quantum growing mode and is thus indispensable. 

\begin{figure*}[t!]
\centering
\subfloat[$n=4$ Garfinkle spacetime.]{%
\includegraphics[width=0.45\textwidth]{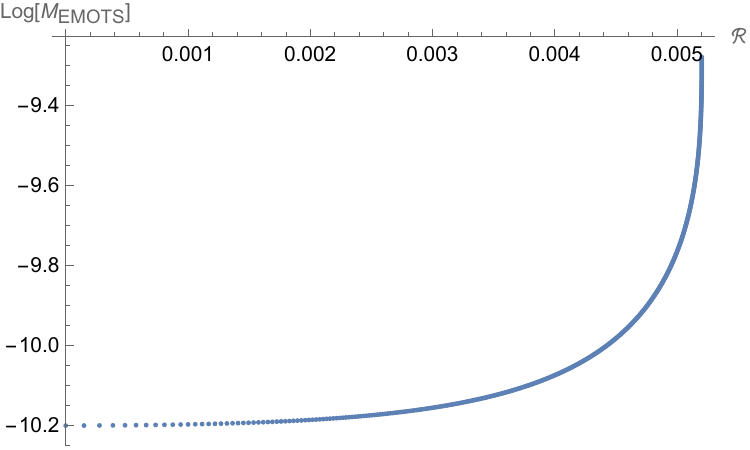}%
  \label{fig:garfinkle-n4}
}
\subfloat[Roberts spacetime.]{%
\includegraphics[width=0.45\textwidth]{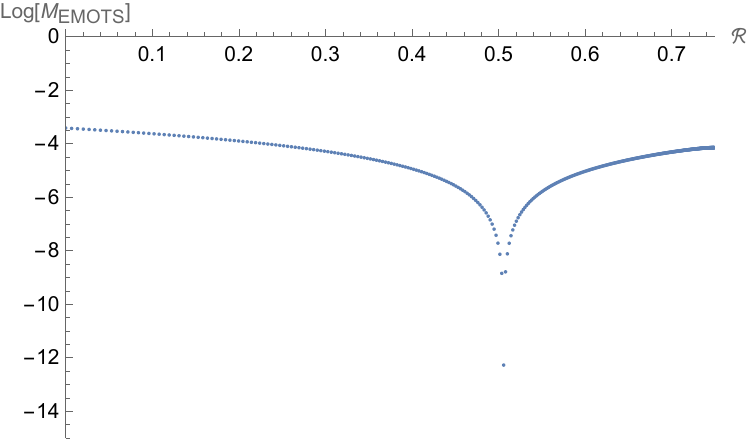}%
  \label{fig:roberts}
}
\caption{(a) For the most physically relevant $n=4$ Garfinkle solution, we plot the logarithm of $M_{\text{EMOTS}}$ against the ratio defined in Eq.~\eqref{eq:ratio}. As $\mathcal{R}$ decreases, effectively moving toward $p \to p^\ast_q$, the mass function monotonically approaches a constant value, indicating a Type I behavior. (b) For the Roberts spacetime, note the Hawking mass in four dimensions is defined as $M \equiv \frac{\bar{r}}{2}[1-(\nabla \bar{r})^2]$. As $\mathcal{R}$ decreases, quantum corrections cause an abrupt, non-monotonic change in the behavior of $M_{\text{EMOTS}}$, likewise signaling a quantum-induced mass-gap transition. }
\label{fig:garf-roberts-wide}
\end{figure*}

A full treatment must also include the classical growing mode $(p-p^\ast)e^{\omega_c T}$, arising from deviations off the critical point $p^\ast$, where the exponent $\omega_c$ for each $n$ has been worked out in~\cite{Garfinkle:2002vn, Jalmuzna:2015hoa}. This allows us to study the competition between quantum vacuum polarization and classical supercritical evolution in shaping the near-critical collapse.

The backreacted geometry then acquires quasi-CSS perturbations of the form
\bea \label{eq:perturbation1}
e^{2 \rho(T,x)}&=&e^{2 \rho_0}+(p-p^\ast) F_c(x) e^{\omega_c T}+\hbar F_q(x) e^{\omega_q T}, \quad
\\
\label{eq:perturbation2}
r(T,x)&=&r_0+(p-p^\ast) r_c(x) e^{\omega_c T}+ \hbar r_q (x) e^{\omega_q T},
\eea
where $\omega_q=1$. For each $n$, the functions $F_i(x)$ and $r_i(x)$ admit closed, real-analytic forms~\cite{Tomasevic:2025clf}. 

To probe horizon formation, we compute the quasi‐local Hawking mass~\cite{Hawking:1968qt, Hayward:1993ph, Misner:1964je, Hernandez:1966zia}
\be \label{eq: Hawkingmass}
M(T,x) \equiv \bar{r}^2- (\nabla \bar{r})^2, \quad \bar{r} =e^{- T} r (T,x).
\ee
Under spherical symmetry, apparent horizons satisfy $(\nabla \bar{r})^2=0$, while $(\nabla \bar{r})^2<0$ identifies trapped regions. The non-linearity of $(\nabla \bar{r})^2$ allows perturbations to reach $O(1)$ in the horizon tracing while remaining within the linear perturbation regime of the background. A careful numerical analysis \cite{Tomasevic:2025clf} reveals a remarkable feature: quantum corrections \textit{lower} the critical point from $p^\ast$ to a new value $p^\ast_q<p^\ast$. Since there are only two dynamically growing modes, it is natural to quantify the competition between the classical and quantum modes by the ratio 
\be \label{eq:ratio}
\mathcal{R} \equiv \frac{e^{\omega_c T }(p-p^\ast_q)}{e^{\omega_q T}  \hbar},
\ee
and evaluate the logarithm of the mass at the earliest marginally outer-trapped surface (EMOTS) where $M_{\text{EMOTS}}=\bar{r}^2(T_{\text{EMOTS}}, x_{\text{EMOTS}})$. As $p \to p^\ast_q$, the would‐be Type II collapse acquires a non-zero black hole mass, indicating \textit{a phase transition to a quantum-induced Type I behavior} with a finite mass gap at the shifted threshold. This is illustrated in Fig.~\ref{fig:garfinkle-n4}.

\paragraph{\textbf{Roberts spacetime in 3+1 dimensions.}} This analytic CSS spacetime is closely related to the DSS solution first identified by Choptuik \cite{Brady:1994aq, Frolov:1997uu, Frolov:1998tq, Frolov:1999fv}. It takes a simple form featuring a null singularity as $T \to \infty$, distinct from the Garfinkle case:
\be
ds^2=2 e^{-2T}e^{2x} [(1-e^{-2x})dT^2-2  dT dx +d \Omega^2],
\ee
with the scalar field $f=\sqrt{2}x$ and $x \in [0,\infty)$. 

We do not identify the Roberts solution with the true $3+1$-dimensional scalar critical solution (which is DSS). Rather, we use it as an analytically tractable CSS proxy whose perturbation structure is closely tied to the breaking of CSS into DSS. In particular, it was found that Roberts exhibits complex growing modes that already encode the oscillatory structure expected when a CSS background is driven toward a DSS attractor~\cite{Frolov:1997uu, Frolov:1998tq}, and previous analyses show that generic growing perturbations evolve away from Roberts toward the DSS critical solution~\cite{Frolov:1999fv}. The role of the Roberts example in the present work is therefore to test whether the semiclassical shielding mechanism survives in a setting analytically connected to the true DSS problem. We return to the true DSS case in Section~\ref{sec:discussion}, where we explain why the same basic semiclassical mechanism is expected to persist there as well.

Regularity again fixes a unique Boulware-like state and the resulting stress tensor behaves as~\cite{Tomasevic:2025clf}
\be \label{eq:Robertstensor}
\langle T^{(4)}_{\mu \nu} \rangle = e^{2T} F_{\mu \nu}(x),
\ee
with a scaling exponent differing from Eq.~\eqref{eq:Garfinkletensor}. A similar backreaction and horizon-tracing analysis reveals a qualitatively similar, but much more exotic transition, as shown in Fig.~\ref{fig:roberts}. The same basic mechanism survives: one-loop vacuum polarization shifts the threshold and replaces the classically naked endpoint by a finite trapped configuration.

\section{Discussion and Outlook} \label{sec:discussion}

\paragraph{\textbf{General lessons.}} The semiclassical framework based on the trace anomaly provides a controlled setting for studying critical collapse in the Einstein--scalar system, even in the presence of non-conformal matter and genuinely time-dependent, self-similar geometries that lack an exact timelike Killing vector. It therefore offers a way to address long-standing questions about quantum effects in critical collapse in a regime where neither static intuition nor conformal simplifications apply. While our analysis uses analytically tractable CSS examples, its main conclusions are not specific to the Garfinkle or Roberts spacetimes; these examples are best viewed as concrete realizations of a more general mechanism.

This mechanism has two key ingredients: the regularity criterion, which selects an asymptotically Minkowskian vacuum-polarization state in the near-critical self-similar spacetime, and the areal-radius scaling, which determines how the reduced stress tensor is lifted to the physical higher-dimensional one. Together they produce a quantum growing mode that competes with the classical unstable mode, shifts the critical threshold, and generates a finite mass gap. The examples studied here are therefore not intended as final models of realistic collapse, but as analytically tractable settings in which the semiclassical shielding mechanism can be exhibited explicitly, even if the detailed coefficients remain model-dependent. We now discuss the broader lessons suggested by this mechanism beyond the particular models considered here.

First, the fact that the regularity criterion singles out a Boulware-like state is mathematically remarkable, since a curved spacetime typically admits infinitely many inequivalent vacua. Physically, however, this is precisely the relevant state for the pre-horizon near-critical regime, since it captures the self-energy of collapsing matter before any horizon forms. At leading order, this conclusion is insensitive to the usual scheme/state ambiguities that often complicate semiclassical gravity in dynamical spacetimes.

Second, although the effective two-dimensional stress tensor is $T$-independent in the reduced description, the physical higher-dimensional stress tensor inherits apparent time dependence through the dilaton-encoded areal radius in Eq.~\eqref{eq:swave}. By self-similarity, this yields a universal growing factor
\be
\langle T^{(D)}_{\mu \nu} \rangle \sim e^{(D-2)T} F_{\mu\nu}(x^i),
\ee
for any spherical self-similar critical spacetime. This growing mode is therefore not an accidental feature of a particular example, but a direct consequence of self-similar geometry itself. Once it is combined with the classical unstable mode, one-loop effects shift the critical threshold, and the would-be naked critical endpoint is replaced by horizon formation and a finite mass gap. In this sense, vacuum polarization dynamically enforces horizon protection even under arbitrary fine-tuning.

This perspective also suggests a broader structural lesson. Since the growing factor is tied to self-similarity rather than to loop counting, higher-loop corrections should inherit the same universal factor while carrying additional powers of $\hbar$. Schematically,
\be \label{eq:higherloop}
\langle T^{(D)}_{\mu \nu} \rangle \propto e^{(D-2)T}
\left(\hbar F_{\mu \nu}(x^i)+\hbar^2 \tilde{F}_{\mu \nu}(x^i)+\cdots \right).
\ee
Because higher loops remain parametrically suppressed, this strengthens the consistency of the linear perturbative treatment in the regime we study, while also suggesting that the non-linear regime---where both the classical and one-loop quantum modes become $O(1)$---may still be accessible within semiclassical gravity.

Finally, these lessons are extracted from CSS critical spacetimes, whereas for $D\geq 4$ the true Einstein-scalar critical solutions are DSS~\cite{Gundlach:2025yjee}. Nevertheless, the same scaling logic should continue to apply: the growth $e^{(D-2)T}$ is then multiplied by a bounded periodic modulation in the echoing variable. Such modulation neither enhances nor diminishes the overall growth, nor does it remove the competition between the classical unstable mode and the quantum growing mode. We therefore expect the threshold shift and the quantum-induced mass-gap mechanism to persist in the DSS case as well.

\paragraph{\textbf{What sets the physical scale of the mass gap?}} In the scale-free Einstein-scalar toy models studied here, the semiclassical mass gap scales parametrically as
\be
M_{\rm gap}\propto \sqrt{\hbar},
\ee
so one might expect
the resulting mass gap to lie near the Planck scale, rendering the semiclassical analysis questionable and seemingly irrelevant for any cosmological applications. However, the key is that the proportionality coefficient is not universal: it depends on the background solution and on the relative strength of the perturbation profiles entering the horizon-tracing condition. Accordingly, the statement that the gap is ``Planckian" should be understood only as a parametric statement within a scale-free model. By itself, the near-critical self-similar region does not determine the absolute physical scale. That scale is fixed only after matching the near-critical patch to a non-scale-free environment through boundary data or through an external scale.

In the $2+1$-dimensional Garfinkle example, this is precisely where $\Lambda$ matters. The shielding mechanism itself does not rely on the cosmological constant at leading self-similar order. But $\Lambda$ enters the quasi-CSS completion that connects the local self-similar interior to the full physical collapse spacetime, and it can therefore affect the physical normalization of both the threshold shift and the mass gap through matching and boundary conditions. This illustrates why the universal mechanism is more robust than the absolute numerical value of $M_{\rm gap}$.

\paragraph{\textbf{Physical implications for primordial black holes.}} Our proposed mechanism may have testable consequences for PBH formation, as critical collapse is seen as one of the main mechanisms for the formation of primordial black holes \cite{Niemeyer:1997mt}. The non-trivial physics of critical collapse comes exactly near the black hole threshold, as self-similarity emerges and a scaling regime takes over. Tuning to this critical regime is quite unphysical if one thinks about astrophysical black hole formation. However, primordial black holes are expected to be formed exactly very close to the black hole threshold. In this cosmological setup, we do not fine-tune the initial data by hand, yet inflation produces a statistical ensemble of density perturbations, and some fraction of these will lie arbitrarily close to the collapse threshold. To a very good approximation, these primordial fluctuations follow a Gaussian distribution,
\be
    P(\delta) = \frac{1}{\sqrt{2\pi} \sigma} \exp\left(-\frac{\delta^2}{2\sigma^2}\right),
\ee
where $\sigma$ is the variance of the primordial power spectrum coming from the inflation model, and $\delta$ is the local density contrast at horizon entry $\delta \equiv \frac{\rho-\bar{\rho}}{\bar{\rho}}$, with $\bar{\rho}$ the mean energy density. The fraction of regions that collapse into black holes is then
\be
    \beta \equiv \frac{\rho_{\text{PBH}}}{\rho_{\text{total}}}= \int_{\delta_c}^1 P(\delta)\, d\delta \propto \exp{\bigg(-\frac{\delta^2_c}{2 \sigma^2} \bigg)},
\ee
with $\sigma \ll \delta_c$, where $\delta_c$ denotes the critical threshold (akin to our $p^\ast$). A well-known drawback of PBH scenarios is that observationally compatible models require delicate fine-tuning of $\delta_c/\sigma$~\cite{Carr:2025kdk}.

Because this probability is exponentially suppressed, the integral is dominated by values just above the threshold $\delta \to \delta_c$. In other words, it is inflationary statistics that naturally selects the near-threshold regime, placing PBH formation squarely within the framework of critical collapse. Moreover, this also implies that the range of masses of PBH that can be formed is not only set by the Hubble scale $M_\text{H}$, it is also controlled by $\delta-\delta_c$,
\be
M_{\text{PBH}}=k M_\text{H} (\delta-\delta_c)^\gamma.
\ee
In fact, for the radiation-dominated era, the cosmological background is well described by a Friedmann-Lemaître-Robertson-Walker (FLRW) spacetime modeled as a perfect fluid, for which a critical solution is known and is precisely of the CSS type~\cite{Evans:1994pj, Maison:1995cc, Neilsen:1998qc}, with $\gamma \simeq 0.36$. This example provides the simplest realization of critical behavior in the cosmological setting. 

Quantum vacuum polarization, which is similarly captured for a radiation fluid by the conformal anomaly, adds a second, quantum growing mode that modifies this picture in two model-robust ways. First, the size of the mass gap shift depends on the relative interplay of the perturbation profiles of the FLRW fluid and quantum backreaction. Nevertheless, it is clear that the mass spectrum will now include a floor value $M_{\text{gap}}$, concentrating PBHs near this gap rather than at an $O(1)$ fraction of $M_\text{H}$~\cite{Tomasevic:2025clf, Niemeyer:1997mt, Yokoyama:1998xd, Green:1999xm, Niemeyer:1999ak, Kuhnel:2015vtw}. Second, the threshold shift can have potentially immense consequences, as even a tiny shift $\Delta \delta$ translates into an exponentially significant change in the abundance of black holes. Therefore, quantum corrections do not merely perturb the classical picture, but can qualitatively reshape both the mass spectrum and the abundance of PBHs.

\paragraph{\textbf{Limitations and outlook.}} Our formalism should be understood as an effective semiclassical description of the dominant $s$-wave sector. Because dimensional reduction and quantization do not commute in general~\cite{Frolov:1999an,Balbinot:2000iy, Cognola:2000wd, Cognola:2000xp}, we do not claim that the reduced two-dimensional renormalized stress tensor must exactly reproduce the full higher-dimensional one under all circumstances. Rather, near criticality, the dynamics is governed by the spherically symmetric sector, while higher-angular-momentum modes are expected to remain subleading under the same regularity and boundary conditions that define the critical solution and its admissible perturbations~\cite{Martin-Garcia:1998zqj}. Within this controlled regime, the reduced dilaton-coupled theory captures the physically relevant vacuum-polarization effect responsible for the quantum growing mode studied here. At the same time, our horizon-tracing analysis is local to the near-critical interior and does not yet provide a complete exterior matching of the quantum corrected spacetime, so a fully global treatment remains an important open problem.

More broadly, an interesting question is whether a soliton phase exists in this quantum-induced Type I collapse. This is motivated by the fact that the threshold shift and mass gap found here closely resemble the phenomenology of classical Type I systems, such as massive-scalar and Einstein-Yang-Mills collapse~\cite{Brady:1997fj,Choptuik:1996yg}, where the introduction of a dynamically relevant scale induces a finite universal mass gap near threshold. In our setup, quantum vacuum polarization furnishes precisely such a scale: $\hbar$ introduces a genuine quantum growing mode that competes with the classical unstable mode until non-linear effects become important. Because vacuum polarization is a generic feature of quantum matter, it is natural to expect that this mechanism may extend beyond the specific models analyzed here, and that Type II collapse may generically cross over to Type I behavior once semiclassical effects are included.

However, unlike classical Type I collapse, where a metastable soliton-like configuration (for example, a boson star or Bartnik-McKinnon solution~\cite{Hawley:2000dt,Bartnik:1988am}) governs the near-threshold dynamics and sets a universal lifetime scaling, our present framework quantizes matter on a fixed self-similar background and treats backreaction only to linear order. Fully establishing whether the semiclassical transition identified here develops into a genuine quantum Type I phase therefore requires going beyond the linear regime and implementing dynamical semiclassical simulations. 

Given the parametric suppression of higher-loop corrections as discussed in~Eq. \eqref{eq:higherloop}, such an extension appears feasible. Encouragingly, recent discovery of soliton-like geometries supported by Boulware-like vacua~\cite{Fabbri:2005zn,Fabbri:2005nt,Ho:2017joh,Ho:2017vgi,Ho:2018fwq,Ho:2019pjr,Barcelo:2019eba,Arrechea:2019jgx,Beltran-Palau:2022nec,Wu:2023uyb,Arrechea:2024cnv,Numajiri:2024qgh} suggests that such a non-perturbative phase may indeed exist. Identifying the relevant metastable semiclassical configuration remains a central open question.

These limitations are especially important for the cosmological implications discussed above. While the qualitative effects of a threshold shift and a finite mass gap are broadly model-robust, a quantitative determination of both $\Delta\delta$ and $M_{\text{gap}}$ for radiation-fluid collapse requires dedicated semiclassical numerical simulations. Only such simulations can determine whether the conjectured consequences for PBHs actually alleviate the tension with observational constraints, and more generally connect the shielding mechanism found here to realistic collapse dynamics. We view this as the important next step in confronting the present framework with observation.

Finally, let us comment on the implications for cosmic censorship. The mechanism identified here is not an ultraviolet resolution of the singularity, but a semiclassical shielding of the classical Type II endpoint. Within this setup, semiclassical matter appears sufficient to protect weak cosmic censorship. It would be interesting to understand whether this mechanism admits a useful holographic interpretation~\cite{Emparan:2021ewh}, starting from states on the holographic side that would evolve into naked singularities. Such states would presumably be non-thermal, at least in the sense that they do not classically correspond to black hole geometries. It is also not obvious whether they admit a standard Euclidean path-integral preparation~\cite{Anous:2016kss}. If such states can be identified, it would be natural to ask what distinguishes them from more familiar semiclassical or thermal states.

At the same time, our treatment neglects higher-curvature effects, which may become important already above the Planck scale. In particular, for small string coupling, there is a hierarchy between $\ell_s$ and $\ell_p$, and stringy physics can intervene before the semiclassical regime reaches its endpoint, potentially replacing the present mechanism by a string-theoretic resolution such as the ``string star" picture~\cite{Horowitz:1996nw, Horowitz:1997jc, Ceplak:2023afb, Seitz:2025wpc, Emparan:2024mbp}. From this perspective, both the semiclassical mechanism studied here and the stringy one suggest that generic naked singularities may be pre-empted already in regimes that remain parametrically far from strong-coupling quantum gravity.

\smallskip

\emph{Acknowledgements.} We thank David Garfinkle, Viqar Husain, Gabor Kunstatter, and Edward Witten for useful discussions. In particular, we thank Roberto Emparan, Carsten Gundlach, and Gustavo J. Turiaci for reading the draft and providing valuable feedback that clarified various subtleties. CHW also thanks Gustavo J. Turiaci for numerous discussions that significantly improved this work. He further acknowledges an audience member at the 39th Pacific Coast Gravity Meeting, whose suggestion inspired this project, and is deeply indebted to Gary Horowitz and Jiuci Xu for their early collaboration and many insightful discussions. CHW is supported by the University of
Washington and DOE Award DE-SC0011637 and DE-SC0026287. MT is supported
by the European Research Council (ERC) under the European Union’s Horizon 2020 research and
innovation programme (grant agreement No 852386). MT is also supported by the Emmy Noether
Fellowship program at the Perimeter Institute for Theoretical Physics.

\bibliography{bibliography}

\end{document}